\documentclass[runningheads,a4paper]{llncs}

\usepackage{amssymb,amsmath}
\usepackage{graphicx}

\usepackage{url}
\urldef{\mailsa}\path|woelki@lusi.uni-sb.de|
\newcommand{\keywords}[1]{\par\addvspace\baselineskip
\noindent\keywordname\enspace\ignorespaces#1}

\begin{document}

\mainmatter  

\title{Phase transitions in cellular automata\newline for cargo transport and kinetically constrained traffic}

\titlerunning{Phase transitions in cellular automata}

%
%
\author{Marko Woelki%
}
\authorrunning{M. Woelki}

\institute{Theoretische Physik, Universit\"at des Saarlandes,\\
66123 Saarbr\"ucken, Germany\\
\mailsa\\
}

%
%

\toctitle{Lecture Notes in Computer Science}
\tocauthor{Authors' Instructions}
\maketitle

\begin{abstract}
A probabilistic cellular automaton for cargo transport is presented
that generalizes the totally asymmetric exclusion
process with a defect from continuous time to parallel dynamics. It appears as an
underlying principle in cellular automata for traffic flow with
non-local jumps for the kinetic constraint to drive as fast as
possible. The exactly solvable model shows a discontinuous phase
transition between two regions with different cargo
velocities.\keywords{asymmetric exclusion process, matrix-product
state}
\end{abstract}

\section{Introduction}
Non-equilibrium phase transitions can rarely be calculated exactly,
i.e.\ without need of approximations or fits of numerical data. One
paradigmatic system where this is possible is the totally asymmetric
simple exclusion process (TASEP) (see \cite{Derrida_Altenberg} and
references therein). The model is defined on a 1d discrete lattice
with sites being either empty (holes) or occupied by a single
particle. A randomly chosen particle moves to the right at rate $1$
provided that the site is empty. For open boundaries with particle
input at rate $\alpha$ and output at rate $\beta$ this leads to
three different phases: a low-density, a high-density and a
maximum-current phase. For finite systems the process can be solved
exactly by the matrix-product Ansatz (see \cite{Blythe} for a recent
and exhaustive review) and the complete thermodynamic behavior that
is relevant for understanding the phase diagram can be extracted
from the asymptotics of this solution. On the ring the process has a
uniform groundstate \cite{Derrida_Altenberg}. However the presence
of a defect particle leads to a rich phase behavior \cite{Mallick}.
In the defect TASEP usual particles move $10\rightarrow 01$ at rate
$1$, the single defect particle moves itself forward $20\rightarrow
02$ at rate $\alpha$ and can be overtaken by usual particles
$12\rightarrow 21$ with $\beta$ The solution is formally related to
the open-boundary case. There is one shock phase and three phases
where it behaves once like a particle, once like a hole and once
like a second-class particle. The second-class particle case
corresponds to $\alpha=\beta=1$. To the left the second-class
particle looks like a hole (as seen from a particle) and to the
right it looks like a particle (as seen from a hole). This case was
studied in \cite{Derrida_2class} since it can be used to study the
shock in the TASEP on the infinite line with step-initial condition.
The connection is that, due to its very special dynamics, the defect
$2$ samples the random shock position on preferring configurations like $000021111$.\hfill\\
The defect TASEP can alternatively be understood as a cargo
transport process: The defect is a usual particle carrying a cargo
which can be handed over to a particle behind. See \cite{braz} for a 
different definition of cargo in the same context. These
mechanisms play a fundamental role in the biology of intracellular transport. 
Processive motors like kinesin transport cargo over long distances \cite{korn}. The cargo can 
alternatively be interpreted as virus particles that use carrier particles in order to attain the interior of a cell, see
\cite{braz} for further references.\hfill\\
The Nagel-Schreckenberg model \cite{nasch} is often referred to as
the minimal model for one-lane traffic-flow on a freeway.
There it is essential to allow for faster and slower cars to get a
realistic flow-density relation: cars can move up to $v_{max}$ sites
per time step. Further all cars are updated simultaneously according
to a parallel update and move independently with probability $p$.
For $v_{\rm max}=1$ it is equivalent to the TASEP with parallel
dynamics. The steady state on the ring shows nearest-neighbor
correlations and has a simple pair-factorized form
\cite{Schreckenberg95}. For open boundaries the matrix-product
technique could be generalized to obtain the exact steady state
\cite{ERS,degier}. It can be interpreted as a pair-factorized state
as on the ring modulated by a
matrix-product state \cite{woelki_par}.\hfill\\
In this article we introduce a generalization of the cargo-transport
process to discrete time with parallel updating and give its exact
solution. Here we restrict ourselves to light cargo, i.e. the case
where the speed of a particle is not lowered by the presence of
cargo. This appears naturally in the steady state of a traffic
cellular automaton \cite{woelki_par}. 
We will see that non-local jump processes where
particles drive as fast as possible can lead for non-deterministic
hopping to a discontinuous phase transition on the ring.

\section{The Cellular-Automaton Model and its Solution}
\begin{figure}
\centering
\includegraphics[height=4cm]{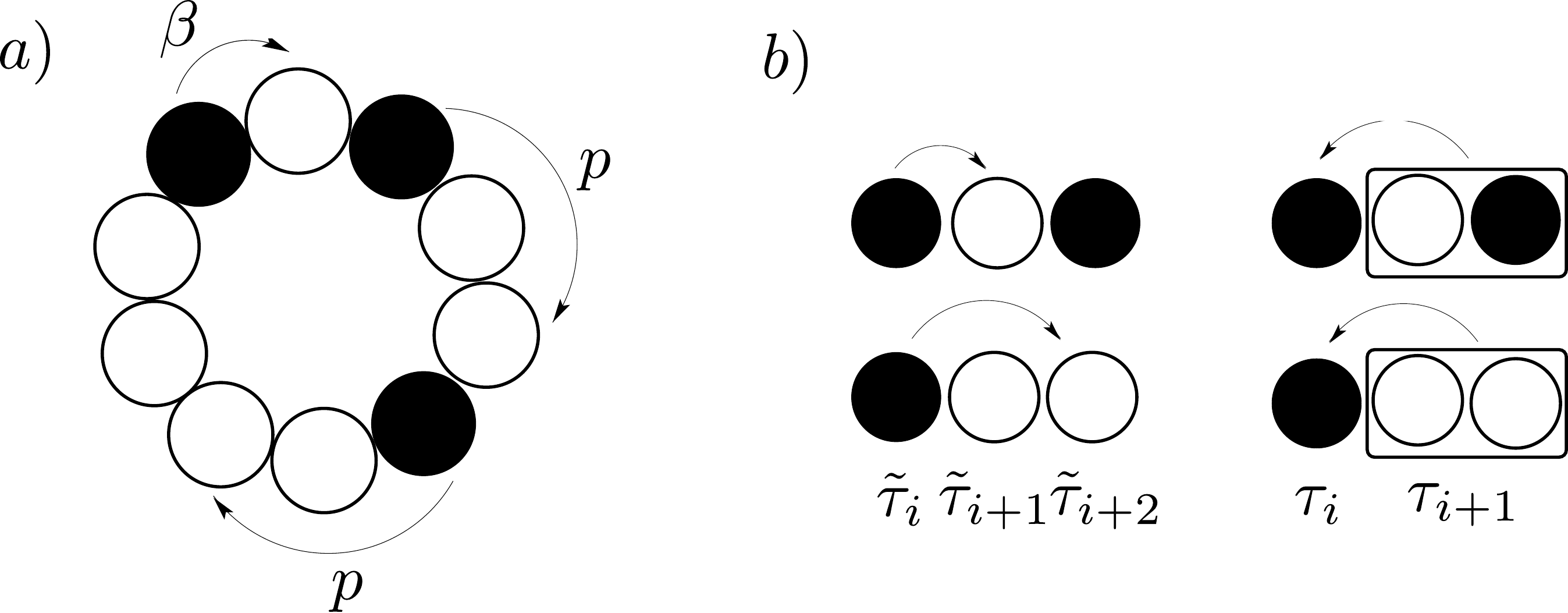}
\caption{a) Allowed moves in a stationary configuration: only one
odd gap between particles b) Equivalence of moves and according
reduction of lattice units} \label{1}
\end{figure}

Consider a periodic one-dimensional lattice with sites being either
occupied by a particle (in state $\tau=1$) or empty ($\tau=0$). One
of the particles carries a cargo to which we refer to as a defect
($\tau=2$). The particles are updated simultaneously and every
particle (with or without cargo) moves forward with probability $p$.
If the site behind it is occupied, the cargo carrying particle can
independently give its cargo back at probability $\beta$. This
simple dynamics is encoded in detail in the transitions
\begin{eqnarray}
\label{rel1}
10 &\rightarrow 01, & \text{at rate } p,\\
020 &\rightarrow x02, & \text{at rate } p,\\
120 &\rightarrow 210, & \text{at rate } \beta(1-p),\\
    &\rightarrow 102, & \text{at rate } (1-\beta)p,\\
    &\rightarrow 201, & \text{at rate } \beta p,\\
\label{rel6}
121 &\rightarrow 21x, & \text{at rate } \beta,
\end{eqnarray}
with $x$ being either $0$ or $1$ indicating that the site can be either empty or occupied due to the parallel update. For example the evolution of the pattern $020$ can be affected by a particle to the left moving itself forward. This is quite a general scenario that 
might apply to biological intracellular cargo transport. The parallel update 
reflects highly active transport where many particles move at the same time.\hfill\\
For $\beta=0$ the cargo is
attached to one special particle for all times and its dynamics is
the same as for the other particles. This corresponds to the usual
TASEP (with parallel update) and the single occupation $\tau=2$ can
be replaced by $\tau=1$. The steady state for a lattice with $L+1$
sites (with one of them occupied by the defect) is
\begin{equation}
\label{P} P(\tau_1,\tau_2,\dots,
\tau_{L+1})=\prod\limits_{i=1}^{L+1}P(\{\tau_{i-1},\tau_{i}\}),
\end{equation}
thus it factorizes \cite{Schreckenberg95} into symmetric two-site
factors $P(\tau_{i-1}\tau_i)\equiv P(\{\tau_{i-1},\tau_i\})$ with
\begin{eqnarray}
\label{P1}
 P(00)&=&1-\rho-J/p,\\
\label{P2}
 P(10)&=&J/p,\\
\label{P3} P(11)&=&\rho-J/p.
\end{eqnarray}
Here $J$ is the particle current
\begin{equation}
\label{J} J(\rho)=\frac{1-\sqrt{1-4p\rho(1-\rho)}}{2}.
\end{equation}
We found \cite{woelki_par} that the steady state for $\beta>0$ can
be calculated exactly too. Here (\ref{P}) is generalized to
\begin{eqnarray}
 P(2,\tau_1,\dots, \tau_{L})&\propto &
\tilde{f}(\tau_1)f(\tau_1\tau_2)\dots f(\tau_{L-1}\tau_L)\tilde{f}(\tau_L)\nonumber\\
&\times&\langle W| \left[ \prod\limits_{i\geq 1} \tau_i
D+(1-\tau_i)E \right]|V\rangle
\end{eqnarray}
This is a pair-factorized state (mainly the steady state for
$\beta=0$) modulated by a matrix-product state. Up to the
normalization this is very related to the TASEP with open boundaries
\cite{ERS,woelki_par}. The vectors $\langle W|$ and $|V\rangle$
represent the defect and the matrices $D$ and $E$ represent
particles and holes respectively. The operators obey the algebra
\begin{eqnarray}
\label{WEE}
\langle W|EE &=& (1-p) \langle W|E,\\
\label{WED}
\langle W|ED &=& (1-p)\left(\langle W|D + p\right),\\
\label{DE}
DE &=& (1-p)\left[D+E+p \right],\\
\label{DV}
D|V\rangle &=& \frac{p(1-\beta)}{\beta}|V\rangle,
\end{eqnarray}
for details see \cite{woelki_par}. Note that the relations
(\ref{DE}) and (\ref{DV}) are quadratic and the other relations are
cubic. Accordingly the dynamical rules in (\ref{rel1}-\ref{rel6})
are quadratic (for $10$ and $12$) and cubic (for $20$) respectively. The thermodynamic particle current (\ref{J}) obviously is unaffected by the cargo.\hfill\\
This steady state appears also in cellular-automaton models for
traffic flow with non-local jumps under kinetic constraint. Consider
the following process on a periodic one-dimensional lattice with
sites being either occupied by one car (in state $\tau=1$) or empty
($\tau=0$). The update rules applied simultaneously to all cars
($\equiv$ particles) are
\begin{eqnarray*}
100 &\rightarrow & 001, \quad {\rm with}\; {\rm probability}\; p,\\
101 &\rightarrow & 01x, \quad {\rm with}\; {\rm probability}\;
\beta,
\end{eqnarray*}
where $x$ denotes either a particle or hole.
The maximum velocity $v_{\rm max}$ thus is two sites per time step
instead of one in the usual TASEP and the kinetic constraint is that
cars can not drive at reduced speed 1 if they could move at maximum
speed. This leads under the parallel dynamics to a non-local
repulsion between cars,
so that finally in the thermodynamic limit only even gaps
($0,2,4,\dots$ holes) have non-vanishing probability. Figure \ref{1}
a) shows schematically the allowed moves in a stationary
configuration.
 For even number of holes the process is equivalent to
the TASEP and for odd number of holes it is equivalent to the
cargo-transport process. In this case a single hole in an
environment of particles and hole pairs is formed that plays the
role of cargo attached to varying particles. In figure \ref{1} b)
one sees that the particle movement of a single site is equivalent
to backward movement of the $01$ position. The $01$ pair plays the
role of the defect, having the characteristic `Janus face', looking
to the left like a hole and to the right like a particle. Usual
holes are replaced by hole twins $00$. To be precise, the
probabilities (\ref{P1},\ref{P2}) would be rewritten here as
$P(00)\equiv P(0000)$, $P(01)\equiv P(001)\equiv P(0001)$. In the
following we restrict ourselves in the terminology to the
cargo-transport process.
\section{The Phase Diagram}
There is a discontinuous phase transition at \cite{woelki_par} the density
\begin{equation}
\label{rhoc}
\rho_{c}=\frac{\beta(1-\beta)}{p-\beta^2}.
\end{equation}
\begin{figure}
\centering
\includegraphics[height=5.2cm]{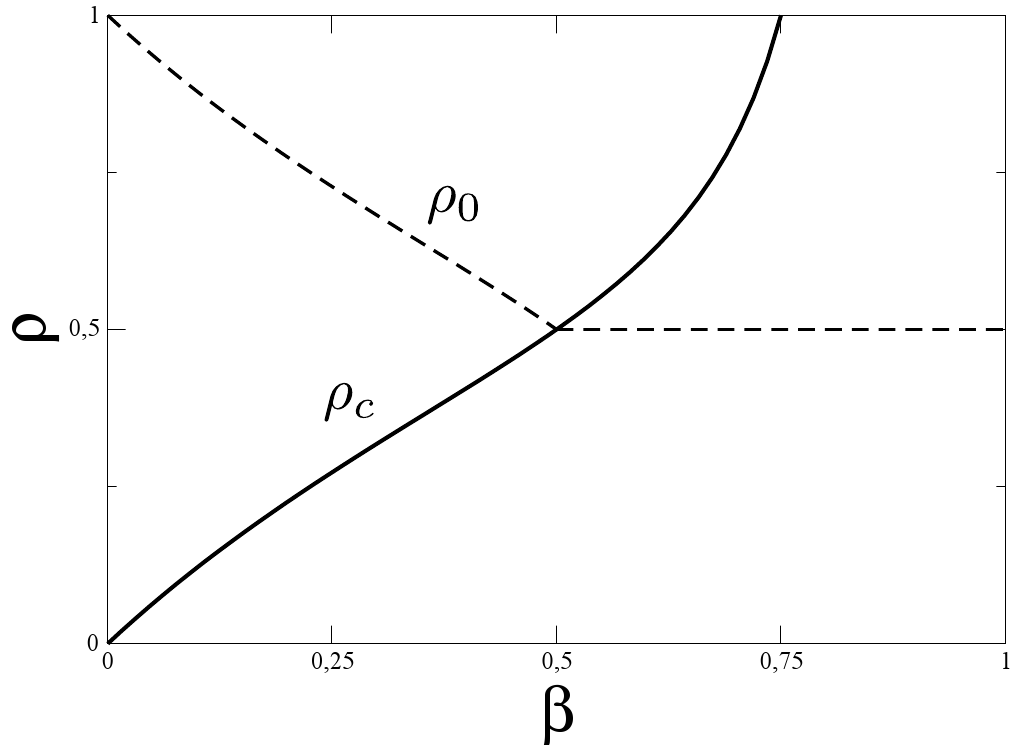}
\caption{The phase diagram shows the $\rho$-$\beta$ plane for
$p=3/4$. The thick line separates the two phases and on the dashed
line the velocity of the defect changes its sign.} \label{phase_dia}
\end{figure}
For $\rho<\rho_c$ and $\rho>\rho_c$ one finds different velocities of 
the defect which can be calculated through
\begin{equation}
\label{v}
v=p(1-\rho_+)(1-\beta\rho_-)-\beta\rho_-.
\end{equation}
Here $\rho_-$ and $\rho_+$ are the densities directly behind and in
front of the defect $2$. The dynamics of the defect is obtained from
(\ref{rel1}-\ref{rel6}). It moves either forward with probability
$p$ if it has a hole in front while at the same time there is no
particle directly behind that simultaneously catches the cargo:
first term in (\ref{v}). Or it moves backwards with probability
$\beta$ if it has a particle behind: second term in (\ref{v}). The
neighboring densities are in terms of the current $J(\rho)$ defined 
in (\ref{J}):
\begin{align}
\label{rhom}
\rho_-=&\begin{cases}
\dfrac{1}{\beta(1-\rho)^2}\dfrac{J^2(\rho-J)^2}{p(\rho-J)^2+(1-p)J^2},
&\rm{ for }\; \rho<\rho_c,\\
\dfrac{p\rho-J}{p\rho-\beta J}&\rm{ for }\; \rho>\rho_c,\end{cases}\hfill\\
\label{rhop}
1-\rho_+=&\begin{cases}
\left( \dfrac{J}{p\rho}\right)^2, &\rm{ for }\; \rho<\rho_c,\\
\dfrac{p-\beta}{p^2(1-\beta)}\dfrac{J}{\rho}& \rm{ for }\; \rho>\rho_c.
\end{cases}
\end{align}
Note that there are many ways to express the results due to the relation $J(1-J)=p\rho(1-\rho)$. 
In other words the square root in $J$ appears in every power of $J$ with a certain prefactor.
Equations (\ref{rhom},\ref{rhop}) yield
\begin{equation}
\label{v2}
\frac{v(\rho)}{p}=\begin{cases}\dfrac{1-2\rho}{1-2J}, &\rm{ for }\; \rho<\rho_c,\\
\dfrac{J-\beta\rho}{p\rho-\beta J} &\rm{ for }\; \rho>\rho_c.
\end{cases}
\end{equation}
The defect velocity vanishes for
\begin{equation}
\label{rho0}
\rho_0=\begin{cases}\dfrac{p-\beta}{p-\beta^2}, &{\rm for }\; \beta<1-\sqrt{1-p},\\
    1/2, &{\rm for }\; \beta>1-\sqrt{1-p},
\end{cases}
\end{equation}
and is positive for $\rho>\rho_0$ and negative for $\rho<\rho_0$.
This leads to the phase diagram, depicted in figure \ref{phase_dia}.
\begin{figure}
\centering
\includegraphics[height=8cm,angle=270]{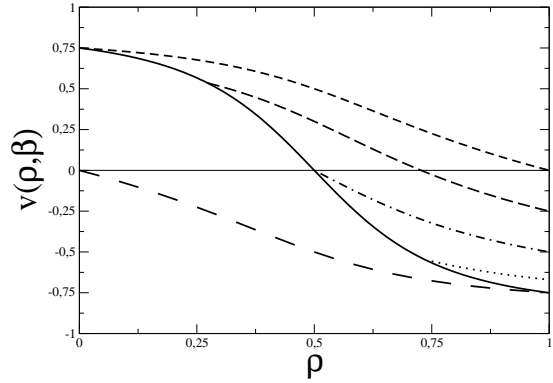}
\caption{Velocity of the defect for $p=3/4$ for varying $\beta$.
{\it Continuous line}: result for $\rho<\rho_c(\beta)$. The {\it
broken lines} correspond to $\beta=0$, $\beta=0.25$, $\beta=0.5$,
$\beta=0.67$ from top ({\it dashed}) to bottom ({\it dotted}) and
are valid for $\rho>\rho_c(\beta)$. The {\it segmented line} is the
velocity of holes} \label{v_dia}
\end{figure}
The formulae (\ref{rhoc}) and (\ref{rho0}) can alternatively be
interpreted in terms of a critical value $\beta_c\equiv
\beta(\rho_c)$ and $\beta_0\equiv\beta(\rho_0)$. This gives
\begin{equation}
 \beta_c=J(\rho_c)/(1-\rho_c)
,\quad \beta_0=J(\rho_c)/\rho_c.
\end{equation}
The value of $\beta_c$ and $\beta_0$ respectively then is
essentially the absolute velocity of holes and particles at the
transition. Figure \ref{v_dia} shows the character of the defect and
the discontinuous phase transition. For $\beta=0$ its velocity is
given by the upper curve and equals the velocity of particles. In
the zero-density limit the velocity is independently of $\beta$
equal to $p$. For increasing $\beta$ the second phase appears: The
velocity of the defect jumps at $\rho=\rho_c(\beta)$ from the
continuous curve to the corresponding dashed curve. $\rho_c$
increases with $\beta$ until $\beta=p$ where $\rho_c=1$, so that the
system is completely in the second phase for all densities. Note
that for $\beta=1/2$ one has $\rho_c=1/2$. Finally in the limit of
the fully occupied lattice $\rho=1$ and $\rho_c\leq 1$ the velocity
equals $\beta$ for $\beta\leq p$. However for $\rho_c>1$ it can
never increase the value of $p$ which there is the corresponding
velocity of the holes given by the segmented line.\hfill\\
For the phase $\rho<\rho_c$, which is purely present for $\beta\geq p$,  
it is important to stress that the quantities $\rho_+$, $\beta\rho_-$ and 
$v$ are independent of $\beta$. The density profile is symmetric around 
the defect and has an algebraic decay. This corresponds to the second-class 
particle phase in the defect TASEP mentioned in the introduction. As in 
continuous time the defect velocity (\ref{v2}) is given by $v=dJ/d\rho$ 
which has the form of a group velocity and becomes $v=1-2\rho$ for small 
$p$, compare table \ref{tab}. Thus the defect travels with the velocity 
of the density disturbance.
\section{Limits}
Table (\ref{tab}) shows the limit of small hopping probabilities:
$p\equiv dt$, $\beta=\tilde{\beta}dt$ with $dt\rightarrow 0$.
Note in comparison that the velocity of normal particles is always
$J/\rho$ which gives $p(1-\rho)+p^2(1-\rho)^2+\dots$ This limit has
been studied in the traffic picture in \cite{woelki_rand,woelki_tgf}
and corresponds to the defect TASEP.
\begin{table}
\begin{center}
\begin{tabular}{c|c|c|c}
$\rho$ & $v$ & $1-\rho_+$ & $\rho_-$\\\hline
$\rho<\rho_c$ & $(1-2\rho)dt$ & $(1-\rho)^2\left[1+2\rho(1-\rho)dt\right]$  & $\rho^2/\tilde{\beta}\left[1-(1-\rho)(1-3\rho)dt\right]$\\
$\rho>\rho_c$ & $(1-\tilde{\beta}-\rho)dt$ & $(1-\tilde{\beta})(1-\rho)\left[1+(\tilde{\beta}+\rho(1-\rho))dt\right]$ & $\rho\left[1-(1-\rho)(1-\tilde{\beta}-\rho)dt\right]$\\
\end{tabular}
\end{center}
\caption{\label{tab} Continuous-time limit $p\equiv dt$,
$\beta\equiv \tilde{\beta}dt$: Values to the order $\mathcal{O}(dt)$
of velocity, empty-space density in front and particle density
behind the defect in the two phases. Here
$\rho_c\sim\beta\left[1-\beta(1-\beta)dt\right]$}
\end{table}
As mentioned before, in the limit $\beta=0$ the defect moves only
forward and loses its role as a defect. Therefore the steady state
is the same as for the TASEP. Thus one has the expressions given in
table (\ref{tab2}).
\begin{table}
\begin{center}
\begin{tabular}{c|c|c}
$v$ & $1-\rho_+$ & $\rho_-$\\\hline
$\; p\dfrac{P(10)}{\rho}=\dfrac{J}{\rho}\; $ & $\; \dfrac{P(10)}{\rho}=\dfrac{J}{p\rho}\; $  & $\; \dfrac{P(11)}{\rho}=1-\dfrac{J}{p\rho}\; $
\end{tabular}
\end{center}
\caption{\label{tab2} Limit of the discrete-time TASEP $\beta=0$}
\end{table}
In comparison the results from (\ref{rhom},\ref{rhop},\ref{v2}) for $\rho>\rho_c$ are rewritten and expanded around the TASEP value:
\begin{eqnarray}
\label{rm}
\rho_-(\beta)&=&\frac{P(11)}{\rho-\beta P(10)}=\frac{P(11)}{\rho}\left[1+\frac{P(10)}{\rho}\beta + \dots\right]\\
\label{rp}
1-\rho_+(\beta)&=&\frac{p-\beta}{p(1-\beta)}\frac{P(10)}{\rho}= \frac{P(10)}{\rho}\left[1-\frac{1-p}{p}\beta-\dots\right]\\
\label{rv}
v(\beta)&=& \frac{J-\beta\rho}{\rho-\beta P(10)}=\frac{J}{\rho} - \left(1-\frac{J^2}{p\rho^2}\right)\beta - \dots
\end{eqnarray}
One sees that $\rho_-$ is mainly the same as for $\beta=0$ but the
density to which the numerator is adressed is reduced by backward
moving so that $\rho_-$ is increased. The same holds for the
velocity $v$. $1-\rho_+$ is even the same as for $\beta=0$ up to a
scale which is, using (\ref{rhoc}) and ({\ref{rho0}), given by  $\rho_c/\rho_0$.\\
For $p=1$ (\ref{rm})-(\ref{rv}) yield the results displayed in table \ref{tab3}.
\begin{table}
\begin{center}
\begin{tabular}{c|c|c|c}
$\rho$ & $v$ & $1-\rho_+$ & $\rho_-$\\\hline
$\rho<1/2$ & $1$ & $1$  & $0$\\
$\rho>1/2$ & $\dfrac{1-\rho-\beta\rho}{\rho-\beta(1-\rho)}$ & $\dfrac{1-\rho}{\rho}$ & $\dfrac{2\rho-1}{\rho-\beta(1-\rho)}$\\
\end{tabular}
\end{center}
\caption{\label{tab3} The partially deterministic case $p=1$: Velocity,
empty-space density in front and particle
density behind the defect.}
\end{table}
For $\rho<1/2$ all particles are separated and move deterministically as in the TASEP. For $\rho>1/2$ the effect of $\beta$ on the velocity remains for all possible values and the phase transition disappears.

\section{Conclusions}
A cellular automaton for cargo transport was introduced that
generalizes the (continuous-time) defect TASEP. The parallel update 
is often more realistic in describing active many-particle transport 
and makes the link between deterministic and random-sequential dynamics. 
The point of interest was a single defect, i.e.\ a particle carrying 
light cargo in  an environment of particles and holes on a periodic 1d lattice. 
Particles move forward with probability $p$ and if a particle is 
directly behind the particle that carries the cargo it may catch the 
cargo with probability $\beta$. 
We found a discontinuous phase transition between two phases with 
different cargo velocities.
Successively increasing $\beta$ lowers its velocity only until $\beta=p$. Then a
saturation effect appears where the velocity becomes independent of
$\beta$. The same holds for the stationary state of a cellular
automaton for traffic where the `cargo' corresponds to small headway
that is formed dynamically. It is attached to one car from behind.
If the subsequent car comes close enough it will catch it up. The
fact that the cargo process appears in a seemingly unrelated
non-local jump process underlines its universal role. The
exact matrix-product state and its cubic algebra holds also for 
the two-species case where multiple cargo is present. It is also trivially generalized 
to case where cargo lowers the speed of the particle to $\alpha<p$.  
For $p=\beta$ and in the presence of several second-class particles it serves 
also as a model system for the study of shocks on the infinite line.
\section*{Acknowledgements}
It is a pleasure to thank Kirone Mallick for his kind
hospitality at IPhT.

\end{document}